\documentclass[twocolumn,aps,prl,showpacs,groupedaddress]{revtex4}
\usepackage{epsfig,amssymb,amsmath,mathrsfs,color}
\usepackage[hypertex,linkcolor=red]{hyperref}

\def\>{\rangle}
\def\<{\langle}
\def\map#1{{\mathscr{#1}}}
\def\N#1{\left|\!\left|#1\right|\!\right|}
\def\Tr{\operatorname{Tr}}
\def\kk{\rangle\!\rangle}
\def\bb{\langle\!\langle}
\def\d{\operatorname{d}}
\def\sH{\mathcal H}

\begin{document}

\title{Memory effects in quantum channel discrimination}

\author{Giulio Chiribella}\affiliation{QUIT - Quantum Information Theory
  Group, Dipartimento di Fisica ``A.  Volta'' Universit\`a di Pavia,
  via A.  Bassi 6, I-27100 Pavia, Italy.}
\author{Giacomo M. D'Ariano}\affiliation{QUIT - Quantum Information Theory
  Group, Dipartimento di Fisica ``A.  Volta'' Universit\`a di Pavia,
  via A.  Bassi 6, I-27100 Pavia, Italy.}
\author{Paolo Perinotti}\affiliation{QUIT - Quantum Information Theory
  Group, Dipartimento di Fisica ``A.  Volta'' Universit\`a di Pavia,
  via A.  Bassi 6, I-27100 Pavia, Italy.}

\date{\today}

\begin{abstract}
  We consider quantum-memory assisted protocols for discriminating
  quantum channels. We show that for optimal discrimination of memory
  channels, memory assisted protocols are needed. This leads to a new
  notion of distance for channels with memory. For optimal
  discrimination and estimation of sets of independent unitary
  channels memory-assisted protocols are not required.
\end{abstract}

\maketitle

\begin{figure}[h]
  \psfig{file=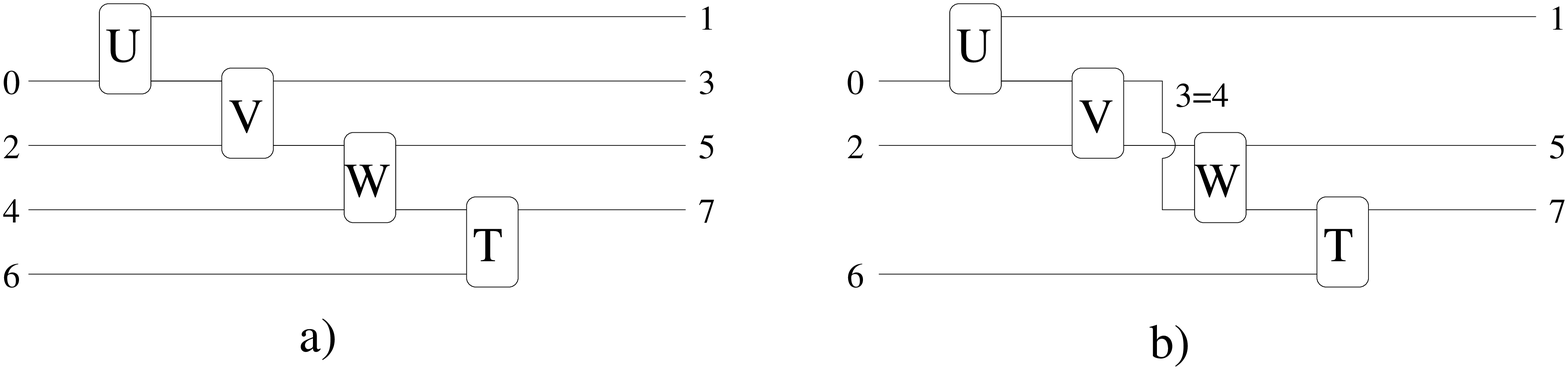,width=8.5cm}
  \caption{\label{uno} Different usage schemes of a general memory
    channel, where the boxes $U,V,W,T$ denote interactions of systems
    with ancillae. a)
    {\em Parallel scheme} (a multipartite input state is evolved
    through the channel). b) A particular case of {\em causal scheme},
    (the output of some use of the channel is fed into a successive
    use).}
\end{figure}

The problem of discrimination between quantum channels has been
recently considered in quantum information
\cite{darloppar,acin,sacchi1,sacchi2,feng,chefles}. For example, in
Ref. \cite{chefles} an application of discrimination of unitary
channels as oracles in quantum algorithms is suggested.  The optimal
discrimination is achieved by applying the unknown channel locally on
some bipartite input state of the system with an ancilla, and then
performing some measurement at the output. A natural extension to
multiple uses is obtained by applying the uses in {\em parallel} to a
global input state. However, more generally, one can apply the uses
partly in parallel and partly in series, even intercalated with other
fixed transformations, as in Ref. \cite{combs}. Indeed, due to its
intrinsic causally ordered structure, the memory channel can be used
either in parallel or in a causal fashion (see Fig. \ref{uno}).  In
this Letter we show that this {\em causal} scheme is necessary when
the multiple uses are correlated---i.~e. for memory channels---whereas
it is not needed for independent uses of unitary channels (the case of
non unitary channels remains an open problem).
Memory channels \cite{palmacc,manc,kretwer,nila,plenio} attracted
increasing attention in the last years. They are quantum channels
whose action on the input state at the $n$-th use can depend on the
previous $n-1$ uses through a quantum ancilla. The problem of optimal
discriminability of two memory channels is relevant for assessing that
a cryptographic protocol is concealing \cite{bitcomm} and for
minimization of oracle calls in quantum algorithms.\par

We will provide an example showing that a pair of memory channels can
be perfectly discriminabile, even though they never provide orthogonal
output states when applied to the same global input state. This new
causal setup provides the most general discrimination scheme for
multiple quantum channels, and this fact leads to a new notion of
distance between channels.\par

In the case of two unitary channels, optimal parallel discrimination
with $N$ uses was derived in Ref. \cite{darloppar,acin}, and in Ref.
\cite{feng} a causal scheme without entanglement was proved to be
equivalently optimal. In the following, we will prove the optimality
of both schemes for discrimination of unitaries. We will generalize
this result to discrimination of sequences of unitaries, and to
estimation with multiple copies. Differently from the case of memory
channels, we will prove that for all these examples causal schemes are
not necessary.\par

It is convenient to represent a channel $\map C$ by means of its Choi
operator $C$ defined as follows
\begin{equation}
  C:=\map C\otimes\map I(|I\kk\bb I|),
\end{equation}
for a channel $\map C$ with input/output states in
$\sH_{\mathrm{in}/\mathrm{out}}$, respectively, where
$|I\kk:=\sum_{n}|n\>|n\>\in\sH_\mathrm{in}^{\otimes2}$, $\{|n\>\}$
being an orthonormal basis for $\sH_\mathrm{in}$. In this
representation complete positivity of $\map C$ is simply $C\geq0$) and
the trace-preserving constraint is
$\Tr_\mathrm{out}[C]=I_\mathrm{in}$.

\begin{figure}[h]
  \psfig{file=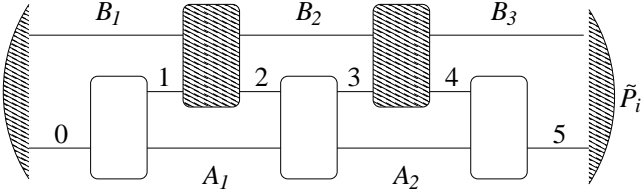,width=8.5cm}
  \caption{\label{genersch} The most general scheme for the connection
    of a memory channel to a quantum circuit corresponding to a {\em
      tester}. The memory channel is represented by its isometric
    gates (white boxes) which denote interaction of quantum systems
    (inputs are labeled by even integers and outputs by odd integers)
    with the ancillae $A_1$ and $A_2$. The tester is represented by
    dashed boxes, including the preparation phase (joint input state
    of system 0 and ancilla $B_1$) and the final measurement stage
    represented by the POVM $\{\tilde P_i\}$.}
\end{figure}

In a memory channel with $N$ inputs and $N$ outputs labeled as in Fig.
\ref{uno}, the causal independence of output $2n+1$ on input $2m$ with
$m>n$ is translated to the following recursive property \cite{combs}
of the Choi operator $C=:C^{(N)}$
\begin{equation}
  \Tr_{2n-1}[C^{(n)}]=I_{2n-2}\otimes C^{(n-1)},\quad\forall 1\leq n\leq N,
\label{caus}
\end{equation}
where conventionally $C^{(0)}=1$. A {\em tester} is a set of positive
operators $P_i\geq0$ such that the probability of outcome $i$ while
testing the channel $\map C$ is provided by the generalized Born rule
\begin{equation}
  p(i|\map C):=\Tr[P_i C].
\end{equation}
The notion of tester is an extension of that of POVM, which describes
the statistics of customary measurements on quantum states. The
normalization of probabilities for testers on memory channels with $N$
input-output systems is equivalent to the following recursive
property, analogous to that in Eq.~\eqref{caus}
\begin{equation}
\begin{split}
  &\sum_iP_i=I_{2N-1}\otimes \Xi^{(N)},\\
  &\Tr_{2n-2}[\Xi^{(n)}]=I_{2n-3}\otimes \Xi^{(n-1)},\quad\forall 2\leq n\leq N,\\
  &\Tr[\Xi^{(1)}]=1.
\end{split}
\label{tester}
\end{equation}
One can prove \cite{combs} that any tester can be realized by a
concrete measurement scheme of the class represented in Fig.
\ref{genersch}.
Mathematical structures analogous to Eqs. \eqref{caus} and
\eqref{tester} have been introduced in Ref. \cite{watgut} to describe
strategies in a quantum game.\par 

Every tester $\{P_i\}$ can be written in terms of a usual POVM
$\{\tilde P_i\}$ as follows
\begin{equation}
P_i=(I\otimes\Xi^{(N)\frac12})\tilde P_i  (I\otimes\Xi^{(N)\frac12}),
\label{povm}
\end{equation}
and for every memory channel $\map C$ the generalized Born rule rewrites as the usual one in terms
of the state
\begin{equation}\label{C}
\tilde C:=(I\otimes\Xi^{(N)\frac12})C (I\otimes\Xi^{(N)\frac12}).
\end{equation}
The state $\tilde C$ corresponds to the output system-ancilla state in
Fig \ref{genersch} after the evolution through all boxes of both the
tester and the memory channel, on which the final POVM $\{\tilde
P_i\}$ is performed \cite{delayed}.\par

The standard discriminability criterion for channels is the following.
Two channels $\map C_0$ and $\map C_1$ on a $d$-dimensional system are
perfectly discriminable if there exists a pure state $|\Psi\kk$ in
dimension $d^2$ such that $\map C_i\otimes\map I(|\Psi\kk\bb \Psi|)$
with $i=0,1$ are orthogonal (every joint mixed state with an ancilla
of any dimension can be purified with an ancilla of dimension $d$).
Here we use the notation $|\Psi\kk:=\sum_{m,n}|m\>|n\>$ which
associates an operator $\Psi$ to a bipartite vector. It is easy to see
that the orthogonality between the two output states is equivalent to
the following condition \cite{noteort}
\begin{equation}
C_0(I\otimes \rho)C_1=0,
\label{discchan}
\end{equation}
where $\rho:=\Psi^*\Psi^T$, where $\Psi^*$ and $\Psi^T$ denote the
complex conjugate and transpose of $\Psi$ in the canonical basis
$\{|n\>\}$, respectively.  The criterion in Eq.~\eqref{discchan} is
too restrictive for memory channels.  Indeed, the correct condition
for perfect discriminability of two memory channels $\map C_i$ with
$i=0,1$ is equivalent to the existence of a tester $\{P_i\}$ with
$i=0,1$, such that
\begin{equation}\label{PC}
\Tr[P_i C_j]=\delta_{ij},
\end{equation}
which means that the two channels can be perfectly discriminated by a measurement scheme as that of Fig. \ref{genersch}.
Using Eqs. (\ref{povm}) and (\ref{C}), Eq. (\ref{PC}) becomes 
$\Tr[\tilde P_i \tilde C_j]=\delta_{ij}$, whence the states $\tilde C_i$ with $i=0,1$ are
orthogonal, and the same derivation as for Eq. (\ref{discchan}) leads to 
\begin{equation}
C_0\left(I\otimes\Xi^{(N)}\right)C_1=0,
\label{condiscr}
\end{equation}
with $\Xi^{(N)}$ as in Eq.~\eqref{tester}.  In Eq.~\eqref{condiscr} the identity operator acts only
on space $2N-1$, differently from Eq.~\eqref{discchan} where it acts on all output spaces.

It is interesting to analyze the special case of memory channels made of sequences of independent
channels $\{\map C_{ij}\}_{1\leq j\leq N}$ and $i=0,1$ (in Fig.  \ref{genersch}, the memory channel
is replaced by an array of channels without the ancillas $A_1$ and $A_2$). The condition for perfect
discriminability is the same as Eq.~\eqref{condiscr} with $C_0$ and $C_1$ replaced by $\bigotimes_j
C_{ij}$ for $i=0,1$, respectively. In terms of a Kraus form $\map C_i=\sum_jK_{ij}\cdot K_{ij}^\dag$
Eq. (\ref{condiscr}) becomes the orthogonality condition $\bb
K_{0j}|\left(I\otimes\Xi^{(N)}\right)|K_{1k}\kk=0$, which for the sequences of maps becomes
\begin{equation}
  \bigotimes_{l=1}^N\bb K^{l}_{0j_l}|\left(I\otimes\Xi^{(N)}\right)\bigotimes_{m=1}^N
|K^{m}_{1k_m}\kk=0.
\end{equation}
for all choices of indices $(j),(k)$, where $K^m_{ij}$ are the Kraus operators for the channel $\map
C_{im}$.  For sets composed by single channels $\map C_i$ with $i=0,1$, the condition becomes simply
the existence of a state $\rho$ such that
\begin{equation}
  \Tr[\rho K^\dag_{0j} K_{1k}]=0,\quad\forall j,k,
\end{equation}
and the minimum rank of such state $\rho$ determines the amount of
entanglement required for discrimination.\par

We now provide an example of memory channels that cannot be
discriminated by a parallel scheme, but can be discriminated with a
tester. Each memory channel has two uses, and is denoted as $\map
C_i=\map W_i\circ\map Z_i$ for $i=0,1$, where the two uses $\map W_i$
and $\map Z_i$ are connected only through the ancilla $A$, and $\map
W_i$ has input $0$ and output $A$ and $1$, and $\map Z_i$ has input
$A$ and $2$ and output $3$. The first use $\map W_0$ of $\map C_0$ is
the channel with $d$-dimensional input and fixed output
\begin{equation}
\map W_0(\rho)=\frac{1}{d^2}\sum_{p,q=0}^{d-1}|p,q\>\<p,q|\otimes |p,q\>\<p,q|,
\end{equation}
$|p,q\>$ being an orthonormal basis in a $d^2$ dimensional Hilbert space. The second use $\map Z_0$
of $\map C_0$ is given by 
\begin{equation}
  \map Z_0(\rho)=\sum_{p,q=1}^{d-1}W_{p,q}\Tr_{A}[\rho(I_2\otimes |p,q\>\<p,q|)]W^\dag_{p,q},
\end{equation}
where the unitaries $W_{p,q}:=Z^p U^q$ are the customary shift-and-multiply operators, with
$Z|n\>=|n+1\>$ and $U|n\>=e^{\frac{2\pi i}d n}|n\>$. The second channel $\map C_1$ is given by
\begin{equation}
  \map W_1(\rho)=\frac I{d^2},\quad\map Z_1(\rho)=|0\>\<0|.
\end{equation}
We will now show that the two channels are discriminable with a casual setup and not with a parallel
one. Their Choi operators are
\begin{equation}
\begin{split}
  C_0&=\frac{1}{d^2}\sum_{p,q=1}^{d-1}|p,q\>\<p,q|_{1}\otimes|W_{p,q}\kk\bb W_{p,q}|_{32}\otimes
  I_0,\\ 
  C_1&=\frac{1}{d^2}\;I^{\otimes 2}_{1}\otimes |0\>\<0|_3\otimes  I_{02},
\end{split}
\label{chois}
\end{equation}
where the output spaces $1,3$ have dimension $d^2$ and $d$, respectively. Suppose that the channels
are perfectly discriminable, then by Eq. (\ref{discchan}) there exists $\rho$ such that
\begin{equation}
  C_0 (I_{13}\otimes\rho_{02}) C_1=C_0 C_1(I_{13}\otimes\rho_{02}) =0,
\end{equation}
where the second equality comes from the expression of $C_1$ in Eq.~\eqref{chois}. Tracing both
sides on the output spaces 1 and 3 one has $\Tr_{13}[C_0C_1]\rho=0$. However,
\begin{equation}
  \Tr_{13}[C_0C_1]=\frac{I}{d^2}
\end{equation}
whence $\rho=0$. This proves by contradiction that the criterion in
Eq.~\eqref{discchan}---corresponding to parallel discrimination
schemes---is not satisfied by channels $\map C_0$ and $\map C_1$. We
will now show a simple causal scheme which allows perfect
discrimination of the same channels.  The first use of the channel is
applied to any state $|\psi\>\<\psi|$, then the measurement with POVM
$\{|p,q\>\<p,q|\}$ is performed at the output on system 1. Depending
on the outcome $\bar p,\bar q$, the second use of the channel is
applied to the state $W^\dag_{\bar p,\bar q}|1\>\<1|W_{\bar p,\bar
  q}$. It is clear that the output of channel $\map Z_0$ is the state
$|1\>\<1|$, whereas the output of $\map Z_1$ is $|0\>\<0|$.\par

This example highlights the need of using a causal scheme in order to
discriminate between memory channels.  The causal discriminability
criterion \eqref{condiscr} implies a notion of distance between memory
channels different from the usual distance between channels. Indeed,
the discriminability criterion (\ref{discchan}) between channels
corresponds to the cb-norm distance \cite{paulsen,kita,notanorm}. The
latter can be rewritten as follows (see e.g. Ref \cite{sacchi1})
\begin{equation}
\begin{split}
&D_{cb}(\map C_0,\map C_1)=\max_{\rho}\N{\left(I\otimes\rho^{\frac12}\right)\Delta
  \left(I\otimes\rho^{\frac12}\right)}_1,\\ &\Delta:=C_0-C_1, 
\end{split}
\label{distcb}
\end{equation}
where the maximum is over all states $\rho$, and $\N{X}_1:=\Tr[\sqrt{X^\dag X}]$ denotes the trace-norm.
One has $D_{cb}(\map C_0,\map C_1)\leq 2$, with the equal sign for perfectly discriminable channels.
For memory channels the discriminability criterion (\ref{discchan}) corresponds to the new distance
\begin{equation}
D(\map C_0,\map C_1):=\max_{\Xi^{(N)}}\N{\left(I\otimes\Xi^{(N)\frac12}\right)\Delta
\left(I\otimes\Xi^{(N)\frac12}\right)}_1, 
\end{equation}
where the maximum is over all $\Xi^{(N)}$ satisfying conditions \eqref{tester}. For $N=1$ this
notion reduces to the usual distance in Eq.~\eqref{distcb}.\par

The easiest application of testers is the discrimination of sequences
of unitary channels $(T_j)$ and $(V_j)$, with $j=1,\dots,N$. Without
loss of generality we can always reduce to the discrimination of the
sequence $(U_j):=(T^\dag_jV_j)$ from the constant sequence $(I)$. Let
us first consider the case of sequences of two unitaries. By referring
to the scheme in Fig. \ref{genersch} we can restate the problem as the
discrimination of $W^\dag(U_1\otimes I)W(U_2\otimes I)$ from $I$ on a
bipartite system, where $W$ describes the interaction with an
ancillary system. It is well known that optimal discriminability of a
unitary $X$ from the identity is related to the angular spread
$\Theta(X)$, defined as the maximum relative phase between two
eigenvalues of $X$ \cite{darloppar,acin}. Apart from the degenerate
case in which $X$ has only two different eigenvalues, the
discriminability of $X$ from $I$ is given by the quantity
$\max\{0,\cos\Theta(X)/2\}\geq0$, which is zero for
$\Theta(X)\geq\pi$, corresponding to perfect discriminability. Since
unitary conjugation preseves $\Theta(X)$ and the angular spread of the
product of two unitaries $X,Y$ satisfies the following bound
\cite{presk}
\begin{equation}
  \Theta(XY)\leq\Theta(X)+\Theta(Y),
\label{spredis}
\end{equation}
and finally $\Theta(X\otimes Y)=\Theta(X)+\Theta(Y)$, one has that
$\Theta[W^\dag(U_1\otimes I)W(U_2\otimes I)]\leq\Theta(U_1\otimes
U_2)$, then no causal scheme can outperform the parallel one. By
induction, one can prove that this is true for sequences of any length
$N$. Indeed, defining $X_{N-1}$ as the product of the tester
unitaries alternated with $U_j\otimes I$ for $1\leq j<N$, if
$\Theta(X_{N-1})=\Theta(\bigotimes_{j=1}^{N-1}U_j)$ holds true, then
it holds also for $N$, due to 
Eq.~\eqref{spredis} . By the same argument, one can also prove that
the sequential scheme of Ref. \cite{feng} equals the performances of
the parallel scheme, since there always exists $T$ such that
$\Theta(UTVT^\dag)=\Theta(U\otimes V)$ (indeed it is sufficient that
$T$ transforms the eigenbasis of $V$ into that of $U$, suitably
matching the eigenvalues).  Therefore, the schemes of Refs.
\cite{darloppar,acin,feng} are optimal also for discriminating
sequences of unitaries. Notice that this also includes the case of
discrimination of two different permutations of a sequence of unitary
transformations.

Another situation in which a parallel scheme already performs optimally is the case of estimation of
unitary transformations $U_g$, $g\in G$ which make a unitary representation of the group $G$.  For
$N$ uses of the unitary $U_g$ the Choi operator in this case is
\begin{equation}
  R_g^{(N)}=R_g^{\otimes N},\; R_g=(U_g\otimes  I)|I\kk\bb I|(U_g^\dag\otimes I).
\end{equation}
The probability density of estimating $h$ for actual element $g$ is
\begin{equation}
  p(h|g)=\Tr[P_h R_g^{(N)}].
\end{equation}
As a figure of merit for estimation one typically considers a cost function $c(h,g)$ averaged on
$h$, with $c(h,g)=c(fh,fg)$ $\forall f\in G$ (the cost depends only on distance, not on specific
location)
\begin{equation}
C_g(p)=\int_G\mu(\d h)c(h,g)p(h|g),
\end{equation}
where $\mu(\d g)$ is the invariant Haar measure on $G$. The optimal density $p$ is the one
minimizing $\hat C(p):=\max_{g\in G}C_g(p)$. For every density $p(h|g)$ there exists a {\em
  covariant} one $p_c(h|g)=p_c(fh|fg)$ $\forall f\in G$ which can be obtained as the average
$p_c(h|g):=\overline{p(fh|fg)}$ over $f\in G$ (practically this corresponds to randomly transforming the
input before measuring and processing the output accordingly).  Since $\hat C(p_c)=\overline{C}(p)\leq\hat C(p)$,
then the optimal density minimizing both costs $\hat C$ and $\overline{C}$ can be chosen as
covariant.  Now, since $p_c(h|g)=p_c(e|gh^{-1})$ ($e$ denoting the identity element in $G$), this
means that the optimal tester must be of the covariant form
\begin{equation}
P_h=(U_h\otimes I)^{\otimes N}P_e(U_h^\dag \otimes I)^{\otimes N}.
\end{equation}
For such $P_h$, the normalization $\int_G\mu(\d
h)P_h=I\otimes\Xi^{(N)}$ implies the commutation $[I\otimes
\Xi^{(N)},(U_h\otimes I)^{\otimes N}]=0$, whence the POVM $\tilde P_h$
in Eq.~\eqref{povm} is itself covariant. The optimal tester problem is
then equivalent to the optimal state estimation in the orbit
$(I\otimes \Xi^{(N)\frac12})R^{(N)}_g(I\otimes \Xi^{(N)\frac12})$.
This proves that the optimal estimation of $U_g$ with $g\in G$ compact
group can be reduced to a covariant state estimation problem, and the
parallel scheme of Ref. \cite{entest} is optimal. The possibility of
achieving the same optimal estimation using a sequential scheme as in
Ref.  \cite{feng} remains an open problem, as, more generally, the
possibility of minimizing the amount of entanglement used by the
tester.\par

In conclusion, we considered the role of memory effects in the
discrimination of memory channels and of customary channels with
multiple uses. We used the new notion of {\em tester} \cite{combs},
which describes any possible scheme with parallel, sequential, and
combined setup of the tested channels.  We provided an example of
discrimination of memory channels which cannot be optimized by a
parallel scheme, and for which the optimal discrimination is achieved
by a sequential scheme. The new testing of memory channels corresponds
to a new notion of distance between channels.  Finally, we showed that
for the purpose of unitary channel discrimination and estimation with
multiple uses, memory effects are not needed.

\acknowledgments This work has been supported by the EC through the
project SECOQC.


\begin{thebibliography}{99}
\bibitem{noteort} The orthogonality condition is
  $0=(I\otimes\Psi^T)C_0(I\otimes\Psi^*\Psi^T)C_1(I\otimes\Psi^*)=H_0^\dag
  H_0 H_1^\dag H_1$, with $H_i=C_i^{\frac{1}{2}}(I\otimes\Psi^*)$,
  which holds iff $H_0 H_1^\dag=0$.
\bibitem{notanorm} This distance is referred to in the literature as
  cb-norm distance, since it is induced by the norm of complete
  boundedness \cite{paulsen} (cb-norm for short, also defined diamond
  norm in Ref. \cite{kita}).
\bibitem{notaspread} In the degenerate case, if $\Theta(U)+\Theta(V)>\pi$,
  it is always possible to find $T$ such that $\Theta(UTVT^\dag)=\pi$.
\end{thebibliography}
\end{document}